\documentclass{article}

\newtheorem{thm}{Theorem}
\newtheorem{defn}{Definition}

\def\refcite#1{\cite{#1}}

\begin{document}

\title{Generalized strong curvature singularities and weak cosmic censorship in cosmological space--times}

\author{Wies{\l}aw Rudnicki\\
Institute of Physics, University of Rzesz\'ow, Rejtana 16A\\
35-959 Rzesz\'ow, Poland\\
rudnicki@univ.rzeszow.pl\\[2ex]
Robert J. Budzy{\'n}ski\\
Department of Physics, Warsaw University, Ho{\.z}a 69\\
00-681 Warsaw, Poland\\
robert.budzynski@fuw.edu.pl\\[2ex]
Witold Kondracki\\
Institute of Mathematics, Polish Academy of Sciences, {\'S}niadeckich
8\\
00-950 Warsaw, Poland}

\date{}

\maketitle

\begin{abstract}
This paper is a further development of the approach to weak cosmic
censorship proposed by the authors in Ref.~\refcite{key-5}. We state and
prove a modified version of that work's main result under significantly
relaxed assumptions on the asymptotic structure of space--time. The
result, which imposes strong constraints on the occurrence of naked
singularities of the strong curvature type, is in particular applicable
to physically realistic cosmological models.\\[2ex]

\end{abstract}

\section{Introduction}

The cosmic censorship hypothesis (CCH) proposed by Penrose \cite{key-1,key-2}
plays a fundamental role in the theory of black holes and is recognized to be one
of the most important open problems in classical general relativity. This hypothesis
states that, in generic situations, all space--time singularities formed in
gravitational collapse, which develops from a regular initial state, should be
invisible to sufficiently distant observers --- by being hidden behind black-hole
event horizons. To be precise, this statement is usually referred to as the
{\it weak} CCH, while its {\it strong} version requires that singularities
resulting from generic gravitational collapse must be invisible to {\it all} possible
observers \cite{key-2}.

The difficulty of the cosmic censorship problem stems from the existence of certain
exact solutions of Einstein's equations which do admit naked ({\it i.e.} not hidden)
singularities. For instance, such sigularities occur in the Tolman--Bondi solution 
representing spherically symmetric inhomogeneuos collapse of dust (see, {\it e.g.},
Ref.~\refcite{key-3}). Naked singularities also occur in more general models of dust
collapse --- namely, in the Szekeres space--times which do not have any Killing 
vectors \cite{key-4}. All these models, however, are highly idealized ({\it e.g.} they 
deal only with the collapse of non-rotating matter), so it is possible that their naked
singularities will disappear under sufficiently generic perturbations. There is,
therefore, the expectation that a suitably refined statement of the CCH may be found,
under which all the possible examples of naked singularities will prove to correspond
to rather exceptional situations. Certainly, such a refinement would involve a precise
criterion of the genericity of space--times and singularities to be considered.

To help identify such a criterion, it may be useful to establish and study various
relations between the curvature strength of singularities and the causal structure
in their neighborhood, as such relations underlie the mechanism of cosmic censorship.
In Ref.~\refcite{key-5} we have defined a new class of singularities, {\it i.e.} the so-called
{\it generalized strong curvature singularities}, that seem to be especially interesting
in this context. Our definition is a generalization of earlier definitions of strong
curvature singularities proposed in Refs.~\refcite{key-6} and \refcite{key-7};
for further motivation for our approach, see the introduction to Ref.~\refcite{key-5}.
As was found in the setting considered in that paper, the assumption that all
singularities are of the generalized strong curvature type cannot be expected to fully
rule out naked singularities --- however, it does significantly constrain the possibility
of their occurrence. Namely, Ref.~\refcite{key-5} proposed and motivated a classification
of such singularities into three types, based on certain relations between the causal
structure of space--time and the focusing properties of solutions to the Raychaudhuri
equation along all causal geodesics in the vicinity of a geodesic that reaches the
singularity. It was then found, under certain additional assumptions imposed on the
asymptotic structure of space--time, that only one of those types of singularities might
evade cosmic censorship.

The main result of the present paper is that a result analogous to Ref.~\refcite{key-5}
can be proven to hold (Theorem~\ref{thm1}) under significantly relaxed asymptotic
assumptions. Namely, Definition~\ref{def1} provides a natural generalization of the
concept of {}``external region'' of black holes to a highly general class of space--times.
This is then employed in the statement of Definition~\ref{def2}, which gives a precise
formulation of the concept of a naked singularity arising from regular initial data.
These assumptions are then found to suffice to prove, as in Ref.~\refcite{key-5}, that
for a wide class of generalized strong curvature singularities (types A and B of the
classification proposed therein), uncensored singularities are ruled out.

Our notation and fundamental definitions will be the same as those in the monograph
of Hawking and Ellis \cite{key-8}.

\section{External Region and Naked Singularities}

In order to define a naked singularity, we first need to give a precise definition
of the external region of black holes occurring in a given space--time. The usual
approaches tend to rely on rather detailed assumptions concerning the asymptotic
structure of space--time. This includes Theorem 1 of Ref.~\refcite{key-5} which applies
to the so-called {\it weakly asymptotically simple and empty}  (WASE) space--times (see p. 225 
of Ref.~\refcite{key-8}). Unfortunately such assumptions, in addition to often being technically 
awkard, may very well be unduly restrictive and physically unrealistic, at least in
a cosmological setting (see Ref.~\refcite{key-9} for a discussion of some problems
with the standard definitions of black-hole region). Considering the current consensus
that the actual universe contains the mysterious dark energy and matter, one would
prefer an approach that reduces assumptions about space--time asymptotic to the extent
possible. Our approach will be based on the following definition which is inspired by
Penrose's concept of {}``external region'' (p. 236 of Ref.~\refcite{key-2}).

\begin{defn}
Let $(M,g)$ be a maximally extended space--time. We denote by $E(M)$ the set of all
points $x\in M$ such that there exists at least one future-complete, achronal null
geodesic with past endpoint at $x$. The interior of $E(M)$ will be called the {\it external
region} of black holes in $(M,g)$.
\label{def1}
\end{defn}

By the requirement that $(M,g)$ be maximally extended we assure the absence of artificial
black-hole regions, {\it i.e.} the regions $M-intE(M)$, which could easily be created even in
Minkowski space by simply removing some 3-dimensional, closed, spacelike subsets.

The region $E(M)$ is defined as the locus of points in space--time from which an
{}``infinitely long'' light ray may be emitted; in other words: that part of space--time
which is visible to some {}``arbitrarily distant'' observers. These light rays are
required to be achronal, in order to rule out the rather pathological situation where
{\it all} future-directed null geodesics are complete without leaving some spatially
bounded region (an example of such a geodesic is that orbiting at radius $3m$ in the
Schwarzschild solution). Moreover, by the same requirement we exclude the possible
presence of causality violations outside the black-hole region. Clearly, this leaves
outside the scope of our considerations the class of naked singularities that are known
to be associated with causality violations \cite{key-10}. However, many familiar examples
of naked singularities, such as those that occur in the mentioned above Tolman--Bondi and
Szekeres solutions, are not associated with causal pathologies.

Let us compare our definition of the external region with that corresponding to the
classical black hole in the maximally extended Schwarzschild solution. As one can easily
see by direct inspection of the Penrose diagram of this solution (see, {\it e.g.}, p. 154 of 
Ref.~\refcite{key-8}), for any point $p$ with the radial coordinate $r>2m$, the achronal boundary,
$\dot{J}^{+}(p,\overline{M})$, of the causal future of $p$  intersects the conformal null
infinity ${\cal J}^{+}$. Fix a point $q\in \dot{J}^{+}(p,\overline{M})\cap {\cal J}^{+}$,  and let
$\mu$ be a null geodesic generator of $\dot{J}^{+}(p,\overline{M})$ outgoing into the past
from $q$.  As $q\in {\cal J}^{+}$, $\mu$ must be future-complete. Moreover $\mu$, when
maximally extended into the past, must intersect $p$, because the Schwarzschild space--time
is globally hyperbolic, and hence causally simple. We thus see that $p$ is a past endpoint of
the future-complete, achronal, null geodesic segment $\mu\cap \dot{J}^{+}(p)$. This means,
according to our definition, that $p\in E(M)$. As $p$ is an arbitrary point with $r$ strictly
greater than $2m$, the set of all such points is open, and so is contained in the interior
of $E(M)$. Since all future-directed null geodesics lying behind the event horizon at $r=2m$ 
will reach the singularity at $r=0$, none of them can be future-complete. Therefore no point with
$r<2m$, {\it i.e.} no point inside the Schwarzschild black hole, can belong to $E(M)$. We thus see
that our definition of the external region coincides exactly with the most standard model of
black hole.

Although we will not do it here, we are able to show that the region $intE(M)$ coincides
exactly with the outer region of the Kerr black hole as well. Moreover, we are able to prove 
a singularity theorem for our black-hole region, {\it i.e.} we can show that if a space--time
$(M,g)$ admits a non-empty region $M-intE(M)$, then there must also exist at least one
incomplete causal geodesic, provided that $(M,g)$ does not contain closed timelike curves
and satisfies the  generic and timelike convergence conditions. As is well known, according
to the celebrated Hawking--Penrose singularity theorem \cite{key-11,key-8}, the same holds
true in the case of any space--time containing a black hole with trapped surfaces.

Observe also that the definition of the region $intE(M)$ involves no assumptions on the
asymptotic structure of space--time --- other than the minimal assumption that future-complete
and achronal null geodesics do exist. Thus $intE(M)$ would appear to provide a natural and
general definition of the external region of black holes in open cosmological models free of
causal pathologies.

We now proceed to formulate a definition of a naked singularity that arises from some regular
initial data given on a partial Cauchy surface $S$. Clearly, the occurrence of a naked
singularity to the future of  $S$ must always lead to the formation of a future Cauchy
horizon $H^{+}(S)$. As usual, a space--time singularity to the future of $S$ is taken to be
implied by the existence of a future-incomplete (null, in this case) geodesic in the causal
future of $S$. By nakedness of the singularity we mean that such an incomplete geodesic is
{}``visible from infinity'' --- {\it i.e.}, remains within the external region $intE(M)$. 
Since the singularity is required to have evolved from some initial data on $S$, we shall assume 
that the intersection of the causal past of the future-incomplete null geodesic with the causal 
future of $S$ will lie within the future Cauchy development, $D^{+}(S)$, of the surface $S$.

\begin{defn}
We say that a {\it naked singularity} arises in the future from regular initial data on a partial
Cauchy surface $S$ when the future Cauchy horizon $H^{+}(S)$ is not empty, and for some point
$r\in H^{+}(S)$ there exists a future-endless, future-incomplete null geodesic
$\lambda\subset J^{+}(S)\cap intE(M)\cap \overline{J^{-}(r)}$ such that

\begin{enumerate}
\item[(i)] $J^{-}(\lambda)\cap J^{+}(S)\subset D^{+}(S)$;
\item[(ii)] $\overline{J^{-}(r)\cap S}$ is compact.
\end{enumerate}
\label{def2}
\end{defn}

The condition of compactness of $\overline{J^{-}(r)\cap S}$ assures the absence of
singularities on $S$ and rules out trivial Cauchy horizons due to a poor choice of the
surface $S$ --- namely, such a choice where some tangents to $S$ tend to null vectors within
a region that is of interest here, a simple example being a spacelike hyperboloid
in Minkowski space.

It should be stressed here that Definition 2 is not particulary restrictive and agrees with other 
definitions of naked singularities considered in the context of the weak CCH. For example, all the 
requirements of Definition~2 are satisfied in the WASE class of nakedly singular space--times, 
for which we have proven our earlier censorship theorem \cite{key-5}. To see this, let us first 
recall that if a WASE space--time $(M,g)$ admits a naked singularity to the future of a partial Cauchy 
surface $S\subset M$, then it cannot be {\it future 
asymptotically predictable} from $S$ (see p. 310 of 
Ref.~\refcite{key-8}). One can then show, under certain reasonable conditions of regularity of 
$S$,\footnote{These conditions require, in essence, that $S$ should have an {\it asymptotically simple 
past} (p. 316 of Ref.~\refcite{key-8}) and that $(M,g)$ should be at least {\it partially} future 
asymptotically predictable from $S$ as defined by Tipler \cite{key-10}. Such requirements are expected 
to be satisfied in any model of gravitational collapse that develops from an initially non-singular 
state.} that there must exist a past-endless, past-incom\-ple\-te null geodesic generator $\eta$ of the 
Cauchy horizon $H^{+}(S)$, which has a future endpoint $p\in {\cal J}^{+}$, and which admits a point 
$r\in \eta\cap M$ such that the set $\overline{J^{-}(r)\cap S}$ is compact; see Lemma~1 of 
Ref.~\refcite{key-5}. Moreover, from condition~(b) of that lemma it is evident that the set 
$I^{-}(p,\overline{M})\cap D^{+}(S)$ must be contained in our external region $intE(M)$.

We shall now show that there must also exist a future-endless, future-incomplete null geodesic
$\lambda\subset J^{+}(S)\cap intE(M)\cap \overline{J^{-}(r)}$, as required in Definition~2.
To this end, let us first define the past set $X\equiv \bigcap_{q\in \eta}I^{-}(q)$.
As $r\in \eta$, we must have $X\subset J^{-}(r)$. Thus, as $\overline{J^{-}(r)\cap S}$ is compact,
$\overline{X\cap S}$ is compact as well. This implies that $X\cap J^{+}(S)$ is a non-empty subset 
of $D^{+}(S)$. In addition, by the definition of $X$, $X\cap J^{+}(S)$ must be a proper subset of 
$D^{+}(S)\cap I^{-}(p,\overline{M})$, because $p$ is a future endpoint of $\eta$ and the closure
of $I^{-}(p,\overline{M})\cap S$ cannot be compact, as $p\in {\cal J}^{+}$. Accordingly,
$X\cap J^{+}(S)$ must be a proper subset of the external region $intE(M)$. Since $\overline{X\cap S}$ 
is compact and $S$ has an asymptotically simple past, by the time-reverse version of Lemma~6.9.3 of 
Ref.~\refcite{key-8}, the past null infinity ${\cal J}^{-}$ must intersect 
$\dot{J}^{-}(\overline{X\cap S},\overline{M})$, where the overdot indicates the boundary in 
$\overline{M}$. We clearly have $J^{-}(X\cap S)=X\cap J^{-}(S)$, so ${\cal J}^{-}$ must intersect 
the achronal boundary $\dot{X}$ as well. Let $\lambda'$ be a null geodesic generator of $\dot{X}$ 
maximally extended into the future from some point in $\dot{X}\cap {\cal J}^{-}$. From the definition 
of $X$ it follows that $\lambda'$ must be future-endless, as the geodesic $\eta$ is past-endless. 
Since $\lambda'$ has a past endpoint on ${\cal J}^{-}$, it must be past-complete; and so it cannot be 
complete in the future. Otherwise, as the generic and null convergence conditions are assumed to be 
satisfied, $\lambda'$ would have to contain, by Proposition~4.4.5 of Ref.~\refcite{key-8}, a pair of 
conjugate points, but this would contradict, by Proposition~4.5.12 of Ref.~\refcite{key-8}, the 
achronality of $\dot{X}$. The asymptotically simple past of $S$ implies that $J^{-}(S)=D^{-}(S)$, 
hence $\lambda'$ must intersect $S$ and enter $J^{+}(S)$.
Let us now denote by $Y$ the set $I^{-}(\lambda')\cap J^{+}(S)$. As $\lambda'\subset \dot{X}$, we
have $Y\subset X\cap J^{+}(S)$. Thus $Y$ must be a proper subset of $D^{+}(S)\cap intE(M)$,
which follows from what has already been established for $X$. There must therefore exist a null
geodesic generator, $\lambda$, of the boundary $\dot{Y}$ contained entirely in $D^{+}(S)\cap intE(M)$.
By the definition of $Y$, $\lambda$ must be future-endless, since the geodesic $\lambda'$ is
future-endless. Moreover, as $\lambda'$ is future-incomplete, by condition~(iii) of Theorem~1 of
Ref.~\refcite{key-5}, $\lambda$ must be future-incomplete as well. From the above proof it is also 
clear that the geodesic $\lambda$ does satisfy all the other requirements of our Definition~2.

In much the same way, one can show that Definition~2 is also satisfied in other classes of 
WASE space--times which are nakedly singular in the sense of weak cosmic censorship --- {\it e.g.} 
in those investigated by Kr\'olak \cite{key-7}. In particular, it is satisfied in the familiar 
Tolman--Bondi solutions with naked singularities \cite{key-3}, because these solutions are WASE 
space--times in which future asymptotic predictability breaks down and they fulfill all the additional 
conditions used in the above proof.  

From the above proof it may also be observed that the breakdown of future asymptotic predictability 
is accompanied by the existence of the two types of incomplete null geodesics in 
$J^{+}(S)\cap J^{-}({\cal J}^{+},\overline{M})$ --- {\it i.e.}, besides the {\it future}-incomplete 
geodesic $\lambda$ there also exists the {\it past}-incomplete geodesic $\eta\subset H^{+}(S)$ which 
has a future endpoint on ${\cal J}^{+}$. This may suggest that our Definition~2 could equally well be 
formulated in a different way, such that some {\it past}-incomplete null geodesic $\eta$ would be 
required to exist in $J^{+}(S)\cap intE(M)$ instead of the {\it future}-incomplete null geodesic 
$\lambda$. Recall, however, that future asymptotic predictability is only a {\it necessary} condition 
on WASE space--times assuring that such spaces are free of naked singularities. Namely, it may happen 
that some WASE space is future asymptotically predictable from a surface $S$ but fails to be 
{\it strongly} future asymptotically predictabe from $S$ as defined by Hawking and Ellis (p. 313 of 
Ref.~\refcite{key-8}).\footnote{In this case one can expect that the smallest perturbation could lead 
to the breakdown of future asymptotic predictability, and so one can require that such cases should not 
occur if cosmic censorship holds in stable, physically realistic space--times.} Then there are no 
singularities to the future of $S$ which are naked, {\it i.e.} which are visible from ${\cal J}^{+}$; 
there must, however, exist some singularities to the future of $S$ which lie on the event horizon 
$\dot{J}^{-}({\cal J}^{+},\overline{M})$, {\it i.e.} some generators of this horizon 
must be incomplete.\footnote{The existence of such generators results from the standard argument with 
conjugate points.} It is not difficult to see that in this case there are {\it no} past-incomplete 
null geodesics in the region $J^{+}(S)\cap J^{-}({\cal J}^{+},\overline{M})$, whereas there 
should still exist in this region some future-incomplete null geodesics that terminate at the 
singularities occurring on the event horizon. In the same way as above, these geodesics can be shown 
to obey all the other requirements of our Definition~2. We can thus say, in accordance with this 
definition, that any WASE space--time, which fails to be strongly future asymptotically predictable from 
a surface $S$, does possess a naked singularity to the future of $S$, which agrees with the widely 
accepted view (see, {\it e.g.}, p. 301 of Ref.~\refcite{key-12}). Clearly, this would not be the case 
if Definition~2 were formulated to require the existence of some {\it past}-incomplete null geodesic in 
$J^{+}(S)\cap intE(M)$ instead of the {\it future}-incomplete one.

\section{Censorship Theorem}

We now turn to the statement of the theorem that forms the main result of the present paper. 
To begin, let us recall that Ref.~\refcite{key-5} introduced the concept of generalized strong 
curvature singularity and proposed a classification of the possible behaviors of geodesic 
congruences in the vicinity of such a singularity (the types A, B and C of Definition~3 in 
Ref.~\refcite{key-5}). One of the results of Ref.~\refcite{key-5} was Theorem~1, stating 
that --- under certain assumptions, including some rather strong assumptions on the asymptotic 
structure of space-time --- the only generalized strong curvature singularities that might evade 
cosmic censorship are those characterized by type C incomplete null geodesics. The theorem we 
state and prove below provides a significant generalization of that result, by doing without any 
detailed assumptions on the asymptotic structure of space--time.

\begin{thm}
Let $(M,g)$ be a space--time admitting a partial Cauchy surface $S$ and satisfying
the null convergence condition, i.e. $R_{ab}K^{a}K^{b}\geq0$ for every null vector
$K^{a}$ of $(M,g)$. Assume also that the following conditions hold:
\begin{enumerate}
\item[(i)] If there exists a future-incomplete null geodesic $\lambda\subset D^{+}(S)$,
then every future-endless null geodesic $\alpha\subset \overline{J^{-}(\lambda)}\cap J^{+}(S)$
is future-incomplete as well (no {}``internal infinity'' in the past of a singularity);
\item[(ii)] Every future-incomplete null geodesic terminates in the future at a generalized
strong curvature singularity and is of type A or B.
\end{enumerate}
Then there exist no naked singularities arising in the future from regular initial
data on $S$.
\label{thm1}
\end{thm}

It should be noted that the assumption of absence of internal infinity, while it holds
for all explicitly known models of gravitational collapse, cannot at present be
guaranteed to hold in the generic case.
\vspace{3ex}

\noindent
\textbf{Proof } The proof proceeds by contradiction.
That is, we assume the existence of a naked
singularity, according to our definition, and demonstrate that this implies that
the assumptions of our theorem cannot hold.

First, we will show that, under the assumption of a naked singularity, there must exist
a future-incomplete null geodesic terminating in a generalized strong curvature
singularity but that is not of type A.

According to Definition 2, there exist a point $r\in H^{+}(S)$ and a future-endless,
future-incomplete null geodesic $\lambda\subset J^{+}(S)\cap intE(M)\cap
\overline{J^{-}(r)}$, such that $J^{-}(\lambda)\cap J^{+}(S)\subset D^{+}(S)$ and
the set $\overline{J^{-}(r)\cap S}$ is compact. Fix such a point $r$, and let $\Lambda(r)$
denote the corresponding family of all null geodesics $\lambda$ with the above-mentioned
properties. Let ${\cal P}(r)$ be the family of all past sets $P\equiv I^{-}(\lambda)$, where
$\lambda\in\Lambda(r)$; and let $\hat{{\cal P}}$ be a maximal chain determined in
${\cal P}(r)$ by the relation of inclusion. Denote now by $P_{0}$ the intersection of all sets 
$P$ belonging to ${\hat{\cal P}}$. Clearly this set is a minimal element of the chain
$\hat{{\cal P}}$. Note also that $P_{0}$ is non-empty. This is because from the
construction of $\hat{{\cal P}}$ it follows that each member of $\hat{{\cal P}}$
intersects the surface $S$ and the closure of this intersection must
be compact, as it is a closed subset of the compact set $\overline{J^{-}(r)\cap S}$.
Thus there exists a corresponding future-endless, future-incomplete null geodesic
$\lambda_{0}\in \Lambda(r)$ such that $P_{0}=I^{-}(\lambda_{0})$.

Let $\{q_{i}\}$ be a sequence of points on $\lambda_{0}$ with no accumulation point on
$\lambda_{0}$ (such a sequence can always be found, as $\lambda_{0}$ has no future
endpoint). Since $\lambda_{0}\subset J^{+}(S)\cap intE(M)$, for each of the points $q_{i}$
there must exist a future-complete, achronal null geodesic $\alpha_{i}$ running from $q_{i}$
to the future. Fix a sequence $\{\alpha_{i}\}$ of such geodesics and extend each of them
maximally into the past. As we have $J^{-}(\lambda_{0})\cap J^{+}(S)\subset D^{+}(S)$,
each of the (past-extended) $\alpha_{i}$ must intersect the partial Cauchy surface $S$.
The sequence $\{ a_{i}\}$ of intersection points $a_{i}\equiv\alpha_{i}\cap S$ must have
some accumulation point $a_{0}\in S$, since all the points $a_{i}$ belong to
$\overline{I^{-}(\lambda_{0})\cap S}$ which is a compact set. It then follows, by Lemma~6.2.1
of Ref.~\refcite{key-8}, that there exists a limit curve, $\alpha_{0}$, of the sequence
$\{\alpha_{i}\}$ with a past endpoint at $a_{0}$. Since all the $\alpha_{i}$ are future-endless
null geodesics, $\alpha_{0}$ must be a future-endless null geodesic as well. Note that
$\alpha_{0}$ must be contained in $\overline{J^{-}(\lambda_{0})}\cap
J^{+}(S)$; otherwise $\alpha_{0}$ would have to intersect the geodesic $\lambda_{0}$ at some
point that would be an accumulation point of the sequence $\{q_{i}\}$, which is impossible
as $\{q_{i}\}$ has no accumulation points on $\lambda_{0}$. Thus, by virtue of condition~(i)
of our Theorem~\ref{thm1}, $\alpha_{0}$ must be future-incomplete. We must also have
$I^{-}(\alpha_{0})=I^{-}(\lambda_{0})$, since $I^{-}(\alpha_{0})$ belongs to the chain 
$\hat{{\cal P}}$, $I^{-}(\alpha_{0})\subset I^{-}(\lambda_{0})$ and $I^{-}(\lambda_{0})$
coincides with $P_{0}$ which is a minimal element of $\hat{{\cal P}}$. As $\alpha_{0}$ is 
future-incomplete, it must terminate in the future at a generalized strong curvature singularity. 
However, $\alpha_{0}$ cannot possibly be of type A: being a limit curve of the sequence $\{\alpha_{i}\}$
whose all elements are future-complete geodesics that leave $\overline{I^{-}(\alpha_{0})}$, it fails 
to obey the defining property of type-A geodesics (see Definition~3 in Ref.~\refcite{key-5}).

As the next step, we shall now show that the same null geodesic $\alpha_{0}$ cannot be of
type B, either. Suppose that $\alpha_{0}$ were of type B. Then, according to the definition
of type B geodesics given in Ref.~\refcite{key-5}, there would exist a geodesic $\tilde{\alpha}$
belonging to the sequence $\{\alpha_{i}\}$ and a point $\tilde{q}\in
\tilde{\alpha}-I^{-}(\alpha_{0})$,  such that the expansion $\theta$ of some congruence of 
future-directed null geodesics outgoing from $\tilde{q}$ and containing $\tilde{\alpha}$ would
become negative somewhere on $\tilde{\alpha}\cap J^{+}(\tilde{q})$. As $\tilde{\alpha}$ is
future-complete and the null convergence condition holds, by Proposition~4.4.4 of Ref.~\refcite{key-8} 
there would then exist some point $x\in \tilde{\alpha}\cap J^{+}(\tilde{q})$
conjugate to $\tilde{q}$ along $\tilde{\alpha}$. Consequently, by Proposition~4.5.12 of 
Ref.~\refcite{key-8}, there would have to exist a timelike curve from $\tilde{q}$ to some point
$y\in \tilde{\alpha}\cap J^{+}(x)$. But this is impossible, because from the construction of
the sequence $\{\alpha_{i}\}$ it follows that the whole geodesic segment
$\tilde{\alpha}\cap J^{+}(\tilde{q})$ must be contained in the achronal part of $\tilde{\alpha}$.
Hence $\alpha_{0}$ fails to be of type B, as well as of type A. Therefore, having shown that
the assumptions of our Theorem~\ref{thm1} fail to hold when its claim is negated,
the theorem is proved.

\section{Concluding Remarks}

While the result of the present work certainly falls short of resolving
the problem of weak cosmic censorship, we have found significant constraints
on the types of generalized strong curvature singularities which might remain
naked. Moreover, we found it possible to extend the result of Ref.~\refcite{key-5}
to a setting where no detailed assumptions concerning the asymptotic
structure of space--time are needed.

A key element of our approach is the definition of the external region
$intE(M)$. It is clear that an approach based on this definition is
problematic in the presence of causality violations in the outer region
of black-hole horizons. We hope to address this issue in the future.

As argued by Penrose \cite{key-2}, if cosmic censorship is indeed a physical principle, 
any arguments supporting the validity of the weak version of cosmic censorship should also support 
the validity of the strong version --- as physics is governed by local laws. It would be 
therefore interesting to see whether it is possible to prove that generalized strong 
curvature singularities of type A and B are subject to strong censorship as well. 
This remains an open problem for future work.

\section*{Acknowledgements}

We are greatly indebted to an anonymous referee for several helpful comments.
Two of us (RJB and WK) wish to thank P. S. Joshi for valuable discussions
and hospitality at the Tata Institute of Fundamental Research, where
part of this work was carried out. This work was supported by the
Polish Committee for Scientific Research (KBN) under Grant No. 2~P03B~073~24.

\end{document}